# Knowledge Retention through Low-Tech Knowledge Sharing Channels in Loosely-Coupled Networks: A Human-Based Approach

**Rosemary Van Der Meer**
School of Management, Operations and Marketing
University of Wollongong
New South Wales
Email: rvdmeer@uow.edu.au

**Karlheinz Kautz**
School of Management, Operations and Marketing
University of Wollongong
New South Wales
Email: kautz@uow.edu.au

## Abstract

This study examines a human-based approach for knowledge retention that is evolving through various knowledge sharing channels in a low-technology environment with a strong emphasis on social networks in a loosely-coupled inter-organisational government-industry collaboration focused on regional sustainability. Using social network analysis combined with interview and observational analysis, our results show that a combination of close-knit community ties and group interaction promote the development of strong personal networks that provide continued access to group memory to retain the group's knowledge.

## 1 Introduction

In the last decade, there has been an increase in inter-organisational collaborations, particularly between government and industry (Bakker et al. 2011). Collaborative groups provide many benefits but over time, people move on to other job roles or leave the group/organisation entirely. How to retain the knowledge of these people has been a key component of organisational and knowledge management research (Anand et al. 1998; Bresnan et al. 2003; DeLong 2007; Hedberg 1981; Moreland et al. 1996; Stein 1995; Walsh and Ungson 1991).

The problem for these collaborations is that technology-based solutions such as expert systems and virtual communities and human-based solutions such as exit interviews, proposed for organisational knowledge retention, may not be possible in inter-organisational collaborations (Braga de Vasconcelos and Kimble 2007; Satzinger et al. 1999; Turoff et al. 1993). In the organisational context, these services are provided by the affiliated organisations. In government-industry collaborations, technology, support and funds may not be available due to the loosely-coupled structure or public focus of the alliance (Alexander 1993). The question is how do these government-industry collaborations perform knowledge retention?

This study examines a human-based approach for knowledge retention that is evolving through various knowledge sharing channels in a low-technology environment and a strong emphasis on social networks in a loosely-coupled inter-organisational government-industry collaboration focused on regional sustainability. Using social network analysis combined with interview and observational analysis, our results show that a combination of close-knit community ties and group interaction promote the development of strong personal networks that provide continued access to group memory to retain the group's knowledge.

The remainder of this paper first provides the theoretical background and motivation for the research and then describes the research method. A presentation of the case settings and the findings as well as a discussion follows. The paper finishes with conclusions and outlines future research.

# 2 Theoretical Background

This section introduces the key concepts of inter-organisational collaboration and knowledge retention which build the theoretical background of our investigation.

## 2.1 Inter-organisational Collaboration

It is becoming increasingly difficult for organisations to operate successfully while only maintaining an internal focus. Organisations are looking beyond their own borders to offset external pressures through collaborative benefits and the opportunity to gain access to competencies unavailable within their own organisation (Chesbrough 2006).

Inter-organisational collaborations consist of members from multiple organisations who identify with both their affiliate organisation and the collaborative group (Alexander 1993). The majority of the research has been focused on industry inter-organisational collaborations. These relationships are often tightly defined through contractual boundaries that specify the purpose for collaboration and reduce risks for the organisations involved (Mentzas et al. 2006; Mowery et al. 1996; Sun and Scott 2005). Benefits to the firms involved include shared financial costs and risk, combined complimentary competencies, the exchange of knowledge and research (Tang 2008). There is also opportunity to develop products for niche markets that may not be viable for a single organisation or to increase production opportunities or research (Hatten and Rosenthal 2001; Tang 2008).

However, inter-organisational collaborations are not limited to the industry sector. Government and government-industry collaborations provide the organisations involved with multiple perspectives on boundary spanning issues such as jurisdictional, environmental or social impacts (von Malmborg 2003). These collaborations also benefit from access to competencies and knowledge that may not be available internally, the inclusion of industry funds to support social projects, development of social networks and innovative problem solving (Choi and Levine 2004; Lozano 2007; O'Leary et al. 2010; von Malmborg 2003).

The recent rise in inter-government and government-industry collaborations is due to the increase in complex problems with overlapping boundaries, shared responsibilities and need for finances greater than public funds can provide (Jones and Lichtenstein 2008; Kaiser 2011). Examples of these problems include disasters such as the Deepwater Horizon oil leak that involved multiple government agencies and industries across state borders, sustainable development initiatives such as the Monroe 2020 project to create a regional geographical information system or development of community mental health systems involving multiple government and non-profit agencies (Manring et al. 2003; Provan and Milward 1995; USCG 2011).

Common characteristics of these inter-organisational collaborations include: an overall purpose and accepted rules for interaction (Majone 1986); loosely-coupled without formalised (legal) boundaries of the relationship amongst participating organisations (Manring et al. 2003; Manring and Pearsall 2004); low autonomy, being held accountable to a governance body or oversight committee (Alexander 1993); ad hoc existence, forming when needed such as for a particular project (Alexander 1993; Manring and Moore 2006); no identifiable location, budget or staff, instead needed services being provided by the affiliated organisations (Alexander 1993); and a facilitator to coordinate group activities (Alexander 1993).

The problem for government-industry collaborations stems from the characteristics that are common in these relationships. These collaborations are more likely to be focused on a particular project, as described above, and thus considered temporary endeavours even if this may span several years. They are less likely to have a specific location or staff assigned for the collaboration, instead utilising resources from those organisations involved such as meeting rooms and technology support. The personnel assigned from the affiliated organisations can also change over time as members move into other job roles (Choi and Levine 2004; O'Leary et al. 2010). This transient nature impacts group knowledge sharing and retention (Gann and Salter 1998; Scarbrough et al. 2004).

## 2.2 Knowledge Retention

Knowledge sharing is the process of transferring or disseminating knowledge so that it can be utilised and applied by an organisation or group (Lichtenstein and Hunter 2006). However, knowledge shared cannot be utilised if it is not retained for later use in some way. Knowledge retention is about keeping the knowledge accessible. Argote et al. (2003) state it "*…involves embedding knowledge in a repository so that it exhibits some persistence over time*" (p572). Knowledge repositories include physical, virtual and mental storage approaches. Keeping access to knowledge within the organisation or group is critical as continually creating new knowledge or recreating lost knowledge is inefficient and can be costly (Marsh and Stock 2003).

Retention of organisational or group knowledge provides a number of advantages, such as: refining core competencies based on experiences (Hedberg 1981); increase learning amongst personnel (Hedberg 1981); increase group and individual autonomy through improved decision making (Churchmann et al. 1957); reduce costs in developing new projects, ideas or products (Walsh and Ungson 1991); and increased, effective functioning and operations (Schatz 1991).

However, knowledge comes in two main forms: 1) explicit, tangible, codified knowledge that can be easily extracted and recorded, such as from documents and procedures (Davenport and Prusak 2000; Kush et al. 2012; Nonaka and Takeuchi 1995); and 2) tacit knowledge that is difficult to codify or record such as personal knowledge gained through experience, knowledge of a group's interactions, external knowledge that a person has access to (Davenport and Prusak 2000; Kush et al. 2012; Nonaka and Takeuchi 1994).

Research has examined a number of solutions, primarily technologically-based, for retaining the explicit and tacit knowledge. Many of these solutions have been successfully implemented in the organisational context, such as virtual communities, use of expert systems and centralised digital repositories (Braga de Vasconcelos and Kimble 2007; Hender et al. 2001; Lindstaedt 1996; Paul et al. 2004; Satzinger et al. 1999; Turoff et al. 1993). However, implementation of technology-based approaches for knowledge retention can be inhibited by limitations and of Information Technology (IT), behavioural factors that influence adoption of technology processes, motivation to contribute and a predilection by staff to use personal networks to seek knowledge (Bresnan et al. 2003).

There has also been research into human-based solutions for the retention of personal knowledge in organisations. Proposals include establishing mentoring systems to allow experienced staff to pass on their knowledge, exit interviews to capture the insights of departing staff or providing rewards to keep knowledgeable staff within the organisation (DeLong 2007; Hewitt 2008; Levy 2011; Miller et al. 2011; Sitlington and Marshall 2011; Teng et al. 2010; Ward and Wooler 2010; Yang et al. 2012). These solutions are internal methods that look at capturing knowledge before personnel leave. Ward and Wooler (2010) thus suggest maintaining relationships with alumni to maintain contact if needed in the future though this relationship method does take time and effort.

While these methods have been successful in the organisational context, there has been little study on how such knowledge retention is performed in the inter-organisational context. Where there is multiple membership of a group, such as in government-industry collaborations, knowledge retention is impacted by the ongoing turnover of members (Kush et al. 2012). The transient aspects of these groups can result in lost knowledge, broken routines, fragmented or diffused memories of group actions and a decrease in successful group decision making (Hollenbeck et al. 1995; Kush et al. 2012; Lewis et al. 2007; Moreland et al. 1996).

The limited resources available to government-industry collaborations in terms of staff, technology and location space makes implementing some of the technology-based approaches proposed for the organisational context problematic. They may not have the training, staff, skills or technology to implement a document repository or expert system.

Human-based methods of knowledge retention may have more success. Fostering mentoring systems or maintaining relationships through informal networks with members once they have moved on may provide continued access to knowledge. Informal networks provide opportunities to supplement other formal interactions (Alexander 1993; Katz and Kahn 1966). Regular, formal interactions, such as those held as part of an inter-organisational collaboration, provide a mechanism for people to develop informal networks with each other (Assimakopoulos and Macdonald 2003). These networks can provide knowledge sharing and retention benefits where individuals use their informal networks for information or advice (Sitlington 2012).

Inter-organisational collaborations provide members with formal interactions at meetings that aid in the promotion of informal networks. Where use of technology-based knowledge retention methods is limited, informal networks may provide the best knowledge retention processes.

# 3 Research Method

To develop an understanding and theoretical explanation if and how informal networks in loosely-coupled government-industry collaborations retain knowledge in a low-tech collaborative environment an in-depth case study was conducted (Yin 2009). The case study concerned a regional, government-industry collaboration that focused on sustainability issues. The group involved a mix of members representing state and local government agencies, industry, non-profit and educational institutions within the region. The group has been operational since 2002, meets on a regular schedule and has a designated facilitator to manage the group's interactions. Table 1 provides a summary of the case study group. The group provides an opportunity to observe established government-industry collaboration with developed personal networks to analyse how they contribute to the group's knowledge retention (Yin 2009).

| Group | Established | Membership | Member Type | Interaction Frequency | Leadership | Governance |
|---|---|---|---|---|---|---|
| EnviroAlliance (EA) | 2002 | 30 | Local/State Govt, Industry, Education, Non-profit | Bi-monthly + events | Chair | Board |

*Table 1 Case Study Characteristics*

The main data collection methods involved observation of the group interactions, interviews and questionnaires. During a 14 month period, the group was observed at 13 events (group meetings and knowledge sharing sessions). Questionnaires were provided to 23 participants with 20 returned. The questionnaires addressed the development of personal networks and informal knowledge sharing channels by members in addition to demographic information on the participants such as organisational type and role within the network. This data was directional, indicating who members seek out, rather than the assumption of reciprocal communication, to highlight knowledge experts (Wasserman and Faust 1994). Semi-structured interviews were conducted with 9 members from the group including interviews with the Chair (group facilitator) and Board representative and members from each of the main organisational types. Audio-recorded interviews lasted between 30-60 minutes and were transcribed. There was also a review of group documentation and email interaction over the observation period.

Data analysis involved Social Network Analysis (SNA) of the questionnaire data and coding of the interview and observation data. Questionnaires and observed interactions were analysed through SNA using UCINET and NetDraw to develop network maps of the interactions between members to understand development of personal networks and informal knowledge sharing channels (Borgatti 2002; Borgatti et al. 2002; Hanneman and Riddle 2005; Wasserman and Faust 1994). Centrality analysis was applied to examine the degree of contact a member has with others in a network (Wasserman and Faust 1994). Through centrality analysis, network maps highlighted prominent members sought for their knowledge. Network maps also highlighted informal links between members such as social, work and personal connections.

Seed coding from the theoretical framework (Holsapple and Joshi 2000) provided for top-down micro-analysis of the interview and observation data to identify channels of sharing and the methods of interaction in the group knowledge sharing activities (Miles and Huberman 1994). Coding analysis was independently verified by two additional researchers and interpretation of the subsequent coding discussed to develop and refine consensus of the results (Miles and Huberman 1994).

# 4 Case Study Setting

The case under investigation is an environmental and sustainability group for a regional, local government alliance (EnviroAlliance) in Australia. In 2002, five local governments formed an alliance to develop a vision for the region and to explore boundary spanning issues that affect the various governments. The purpose behind the formation was to form a collaborative voice at all levels of government, promote discussion on 'big picture' regional issues and improve multi-agency collaboration and knowledge sharing.

The alliance structure includes an oversight Board of Directors and a set of focused groups that examine key regional issues, one of which is the environment and sustainability group. The Board consists of elected representatives from each of the five local governments. The focus groups are volunteers from state government agencies, the local governments, education, industry and non-profit organisations in the region. Each focus group has a leader to facilitate the group's operations and meeting schedule. Organisations choose whether to participate in one or more focus groups with no formal requirements.

While the local governments involved provide funds for a centralised administrative team for the alliance, the other organisations pay a nominal joining fee. This means that focus groups, such as EnviroAlliance, have no operational budget.

EnviroAlliance's purpose is to provide environmental planning for the region and ensure that regional sustainability matters are at the forefront of the state and local governments' considerations. It is a loosely-coupled group of organisations with an interest in the environment and sustainability.

The organisations involved in EnvironAlliance include the local governments that form the alliance, state and federal government agencies such as the Environmental Protection Agency (EPA), local catchment and utility organisations such as water authorities. Non-government participating organisations include educational institutions, industry, local Small-to-Medium Enterprises (SME) and non-profit organisations. The participants consist of representative members who fall into one or more of the following categories: volunteers to represent their organisation because of the organisations environmental interests; volunteers because of their own, personal interest in sustainability; or attendance is part of their job role as they are their organisation's sustainability officer. Members are co-located within the region providing opportunity for face-to-face meetings.

The members are mostly in the 40-50 age bracket, well educated, articulate and have some interest in sustainability. Most members hold some position of authority in their organisation such as team leader or manager. The majority of members are the sole representative at EnviroAlliance for their organisation. A few organisations share the role of representative at EnviroAlliance between two or three people. This means at most meetings, the same representatives attend. Meetings are held bi-monthly.

Over the course of the investigation, the membership of the group increased from 20 members to 30. Membership not only increased, but evolved with some original members being replaced on the group as the representative for their organisation. As members were promoted or moved on to new positions, a new representative from their organisation would join the group.

# 5 Findings

The group's main forms of knowledge sharing are through the bi-monthly group meetings along with communication outside of the regular meetings. Members also participate in ad hoc meetings for the smaller project groups and through social interactions such as regional environment and sustainability events. Results outlined here describe the human- and technology-based approaches used in the group to retain knowledge. The group has three channels they use for sharing knowledge, the limited application of technology within the group and the importance of the personal networks that have developed as a method of retaining access to knowledge, in particular knowledge of past members.

## 5.1 Channels of Sharing

Members of EnviroAlliance communicated and shared knowledge through group meetings, smaller project specific teams and through the personal networks that the members developed.

The primary channel of sharing was through the group meetings held bi-monthly. These meetings were predominantly high-level face-to-face interactions providing members with the opportunity to develop a broad knowledge of the issues effecting all member organisations and to ensure everyone was kept up-to-date on any project work. They also provided opportunity for group decision making. This was indicated through statements from members such as "*...it's not like I'm learning a lot of things from scratch but there might be just little bits of information that come up that just further develop your understanding of a topic you already know quite a bit about*" (Paul_EA).

Smaller groups based on particular projects were also formed by members. Members volunteered to participate in these project groups based on their own interests. These communications were through ad hoc, face-to-face meetings and via email or phone. At the regular EnviroAlliance meetings, each project group would provide reports on their progress to keep all members up-to-date. Projects undertaken in these smaller groups included developing white papers on new legislation, grant applications and identifying solutions to regional environmental issues.

Knowledge shared between members of the project groups was more in-depth than achieved through the bi-monthly EnviroAlliance meetings. It enabled them to "*...have that creativity, innovation and conversation*" that was not possible in the bi-monthly meetings with the whole group (Hugo_EA). Knowledge communicated in the project groups involved notes and emails between the participating members and the report given at the bi-monthly meetings. No other records of their interactions were maintained.

The third channel of knowledge sharing between members of EnviroAlliance was through the personal networks members developed as individuals. These connections were primarily established through personal, work or social interactions. However, connections had also been developed between members through participating in the EnviroAlliance meetings and project groups. Personal network interaction was facilitated by the regular group meetings as small clusters of members would gather to discuss further issues. This was observed at all group meetings with the small clusters changing as members left or moved to speak to someone else. Members also indicated that they maintained regular face-to-face, email and phone contact with members of their personal networks. These interactions provided opportunity for: discussing in-depth issues that had developed from group meetings as an opportunity to ask *"...nitty-gritty type questions*" (Tim_EA); seeking knowledge from a known expert in the group on a member's own work-related issues such as *"...spontaneous...getting information that you're unaware of that can help you to do better work*" (Paul_EA); and development of collaborations on common issues that have developed from work rolls or group issues *"...a lot of the time it's around you know, your networks that perhaps identify the opportunity*" (Hugo_EA).

EnviroAlliance is reliant on channels of sharing that follow the methods of face-to-face meetings, phone and email. Each channel they have provides different levels of knowledge sharing for members from high level through to context specific depending on the channel. These different channels allow members to develop an overall knowledge of all group activities but also allow for in-depth knowledge sharing tailored to individual needs.

## 5.2 Low Technology-based Retention Approaches

Observation of the group's knowledge sharing over the 14 months identified that there were no technology-based retention methods such as a repository for the minutes from group meetings or decision making support documents. Additional communications, such as the distribution of the minutes or update information, was provided to members via email in between the group meetings.

The group had an outdated website at the beginning of the observation period that was not updated over the subsequent 14 months. When the group Chair (Ethan_EA) was asked about the lack of updates to the website, he indicated that he and other members lacked the skills to maintain the site. The site had originally been developed by the small administrative group established as part of the Alliance but they did not provide ongoing support.

There was very little financial backing for or direct technological support provided to the group. This meant that, as with the website, only the members themselves were available to provide technology support if inclined. When asked about utilising other technologies to promote group communication and maintain a record, such as general web-based or social media-based solutions, members were non-committal, stating they felt it was "*...a lot of noise*" (Matt_EA). When questioning members about their use of technology, most members indicated that they used email and phone primarily for communication. However, they acknowledged the web as a useful tool for sourcing journals and information needed for their work. Beyond that, technology was not considered by members and they seemed generally noncommittal in options to use technology to record and retain the shared knowledge.

Through the use of email to distribute documents and with no repository that members could access as the group grew, new members especially were disadvantage in developing an understanding of the group's previous decisions and discussions. Members replacing a previous representative for their organisation were also affected where the original member did not provide any background information. Thus the lack of technology-based repositories for the group's information did impact on the ability of new member's to quickly become an effective part of the group cohesion. The low-tech knowledge retention approaches were however, outweighed by strong human-based retention approaches as described in the next sub-section.

## 5.3 Strong Human-based Retention Approaches

Through observations of the group's activities, the primary interaction appeared to be the bi-monthly meetings. However, a notable level of personal network development and crucial networks amongst members emerged through further analysis. Our examination of the social networks disclosed a close-knit group with strong personal networks between members. These networks were developed through personal, social and work connections that were also facilitated through participation in the group. Members often formed small, ad hoc discussions after group meetings. The regular interaction between members through the group meetings provided another approach for network development. Figure 1 outlines the network connections members indicated they had developed within the group and the method in which these connections had been formed.

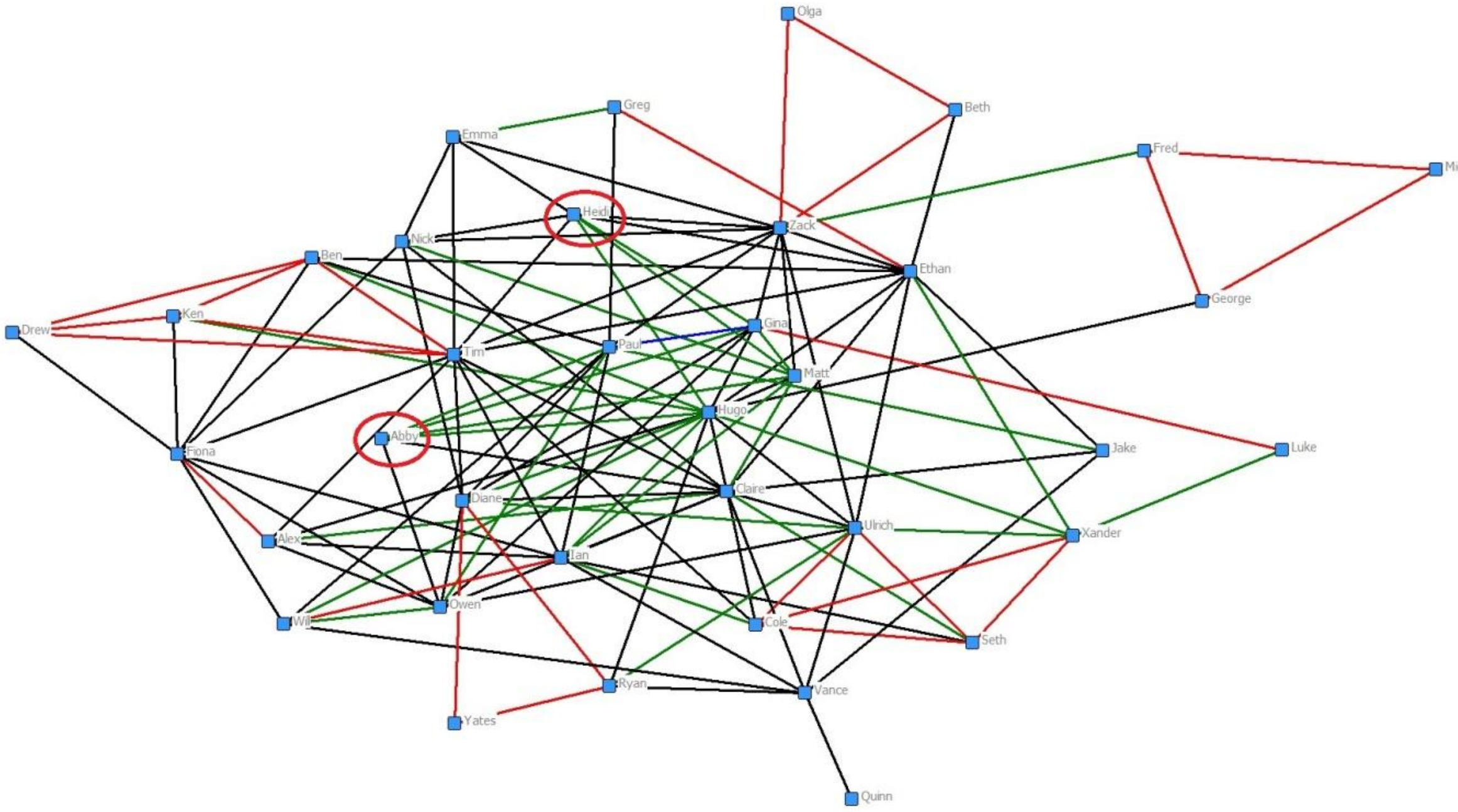

*Figure 1 EnviroAlliance Personal Networks*

Note: the blue line indicates a personal relationship between members (in the centre of the map); red lines indicate work relationships; green lines indicate social relationships; black lines indicate network connections formed through group participation. Network layout defined using Netdraws spring-embedded algorithm.

The analysis of these personal network connections revealed several interesting aspects. One close personal relationship was found between two members (blue line in centre of map). These two members were married but each represented different organisations in EnviroAlliance, a state government department and a local government administration that each had different perspectives and agendas within the group.

A number of work connections highlighted by the red lines showed members that represented the same organisation. However, in only one case did all the members of the same organisation share the representative role for their organisation at EnviroAlliance (triangle in far right corner of map). In all other cases, members either represented different departments from their organisation or different job roles (for example an elected local government representative and an administrative representative).

Green lines indicate other social connections between members. Members indicated that some of these social connections were formed through their own job role. The requirements of their job meant attending events that other EnviroAlliance members also regularly attended, for example those representing local governments. Others indicated that their social interactions had formed due to common interests such as sustainability or because of the smaller size of the regional area, for example Matt_EA talking about Heidi_EA, "*…we live nearby which helps. So we'll often catch up for a drink…and talk about work stuff*" and Paul_EA speaking about Abby_EA, "*…she's one of the sustainability sisters with Gina_EA…so yeah we see her occasionally socially*". Finally, the black lines indicated relationships that had developed between members through their involvement in EnviroAlliance.

The close-knit, personal ties between many of EnviroAlliance's members through different connections show the opportunities for knowledge sharing and retention through and within the relationships between these members also outside of the group meetings. This supports personal networks as a key channel for knowledge sharing and means of knowledge retention between members in an inter-organisational collaboration where members can be representative of many, different organisations.

Two members in particular, highlighted with red circles in the figures, demonstrated the value of these personal networks as a method of retaining knowledge in the group. Heidi and Abby were two members that left the group during the observation period for different reasons. Heidi left at the time observations started, to take on a new role at a different organisation. Abby was with EnviroAlliance for the first few meetings but left due to a restructure at her organisation. Both these members featured prominently in network maps, even after their departure. The maps uncovered their importance as contacts within the group for both practical and financial knowledge in sustainability (see Figure 2 and Figure 3).

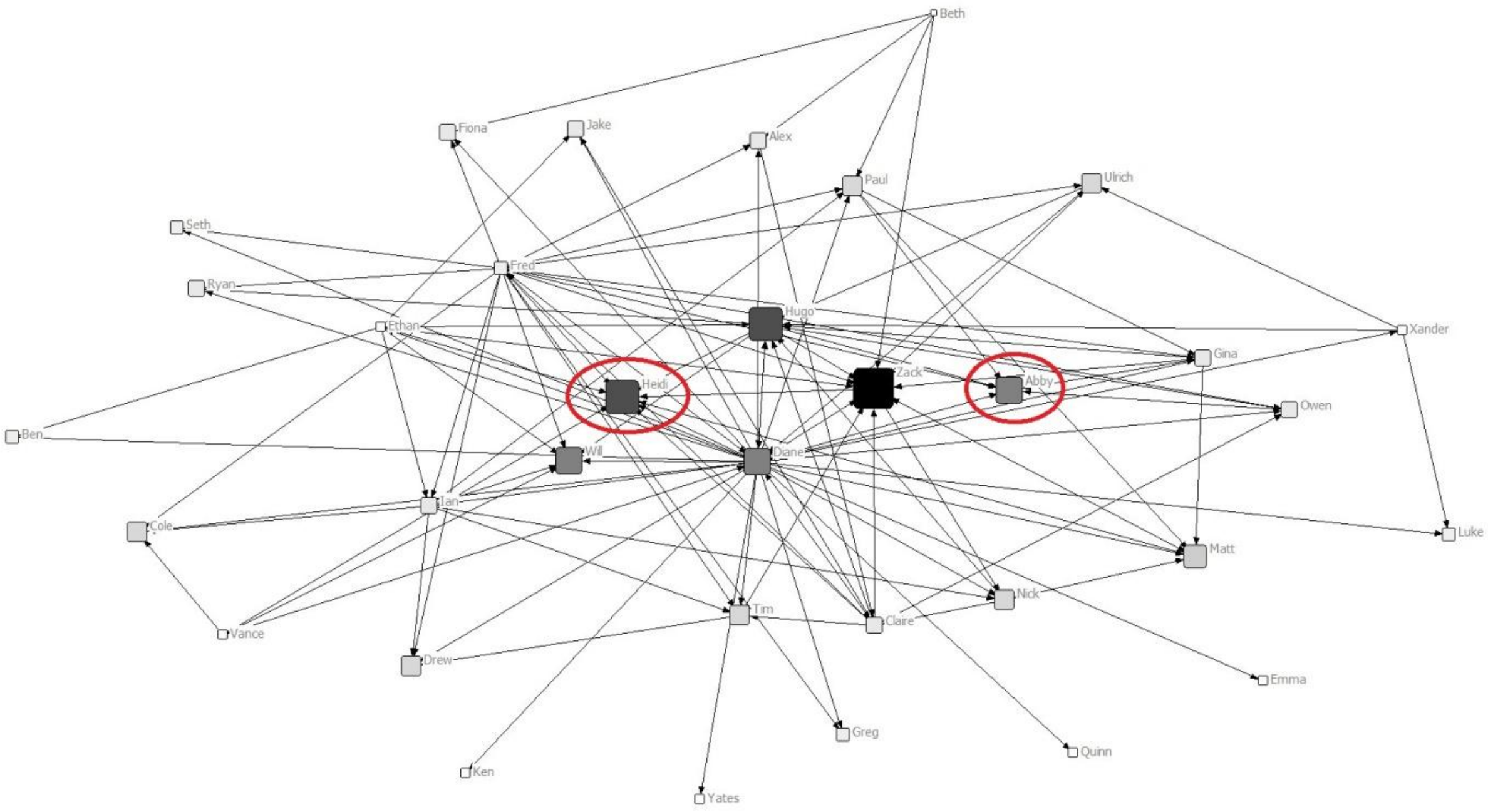

*Figure 2 Network map showing indegree centrality of members for their practical sustainability knowledge.*

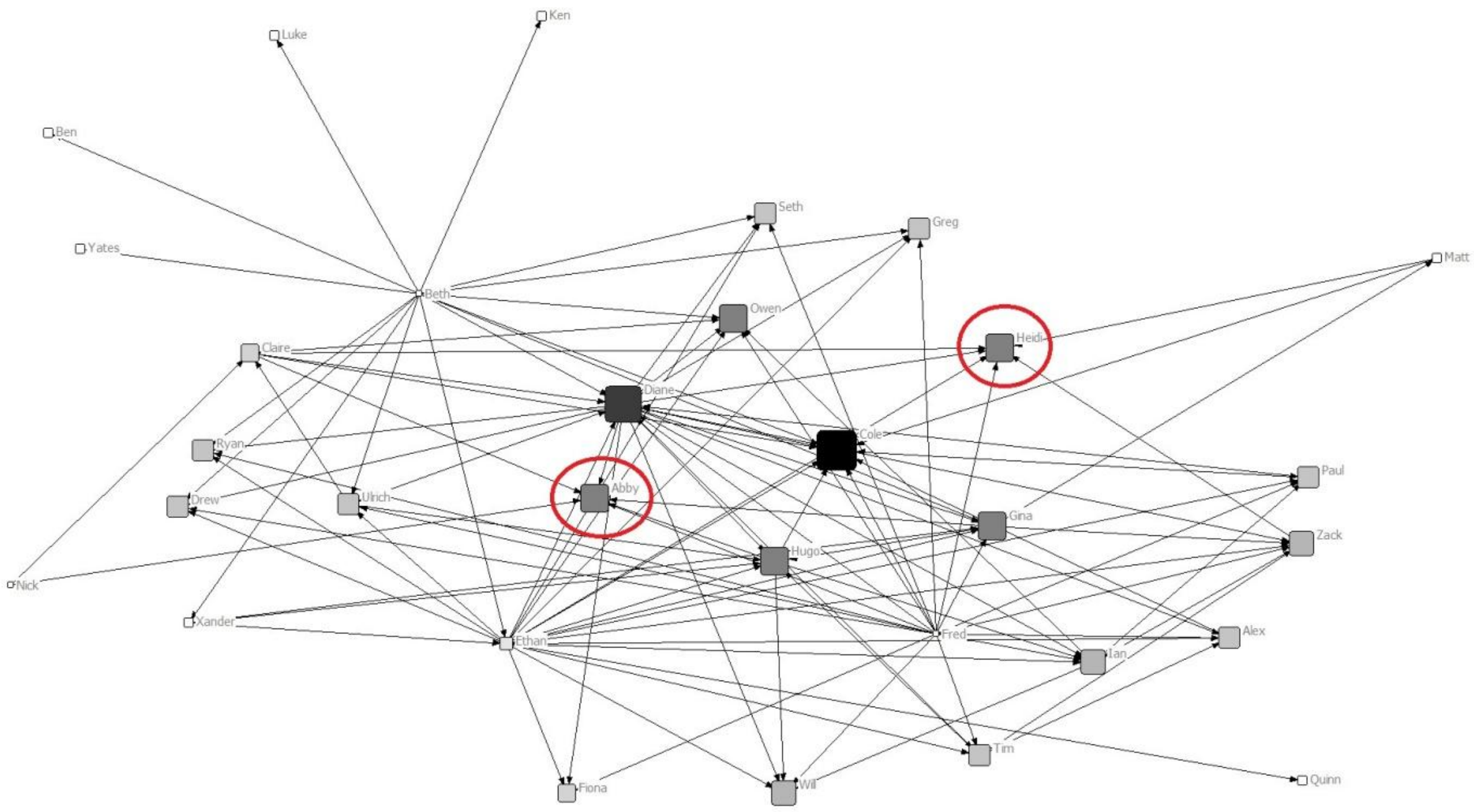

*Figure 3 Network map showing indegree centrality of members for their knowledge of sustainability funding opportunities.*

Note: in Figures 2 and 3, node size and colour determined by indegree centrality, distribution defined by Netdraw's spring-embedded algorithm.

The amount of contact a member receives from other members within the network is indicated by indegree centrality (Wasserman and Faust 1994). While neither Heidi nor Abby have the highest indegree centrality in these maps, they do rate prominently indicating that they are both approached by other members within the network and are recognised as major sources of existing knowledge in practical and financial sustainability issues by EnviroAlliance members (Wasserman and Faust 1994). This perception was supported by members' statements confirming the value of their knowledge such as "(Abby) *will talk about grants that she's got*" (Claire_EA) and "*…she has really good practical skills, …you see that she also lives it as well as talks about it*" (Paul_EA) and for Heidi "*a bit of a sounding board…particularly with a small business built environment sort of angle*" (Matt_EA).

This demonstrates that even though members may leave a collaborative group, their knowledge is retained and through their continuing relationships, not necessarily lost to that group. The social and group network connections formed between members of EnviroAlliance and Heidi and Abby (indicated by the green and black lines in Figure 1) show continued contact through the personal networks. The prominence of both members in the network maps (shown in Figure 2 and Figure 3) evidences the value of their knowledge and their continued participation in the group through the personal networks despite no longer being representatives. The utilisation of these personal networks within EnviroAlliance demonstrates the beneficial application of such a human-based approach to knowledge retention.

# 6 Discussion

The purpose of this study was to examine the knowledge retention approaches in loosely-coupled inter-organisational collaborations. The case examined involves participants from a number of government and industry organisations that volunteer to participate because of an interest in sustainability. The group, EnviroAlliance has no formalised requirements for participation.

We found that the group developed three main knowledge sharing pathways, each with a different level of knowledge and focus. The group meetings provide opportunity to give all members an overview of projects and regional issues. While the knowledge disseminated at this level is high-level, the group meetings also provide opportunity for all the members to interact together. The smaller project groups allow for more specific knowledge sharing on focused topics that is not possible during the group meetings. The personal networks allow for detailed or work –oriented discussion knowledge sharing. These networks are developed through social or work connections but additional contacts are made through the group meetings.

The ongoing interaction amongst members through the group meetings and projects provides a twofold benefit. The meetings provide a path for different levels of knowledge sharing and to ensure all members have a 'big picture' understanding. However, they also supplement the social and work opportunities in network development amongst the members by providing the members with regular interaction (O'Leary et al. 2010).

Our findings show that even though loosely-coupled by organisational representation, the group has a close-knit personal network development. These personal networks provide a pathway for both group and personal knowledge sharing and the retention of, in particular, knowledge of valued members who had moved on for whatever reason. This contrasts Moreland et al. (1996) who found that knowledge retention was greater in intact groups than in those with membership changeover. We demonstrate that where there is a strong personal network development, as has occurred in EnviroAlliance, there is the ability to retain knowledge within the group even when members do leave. EnviroAlliance, through personal networks, has developed a high relative capacity to maintain access to external knowledge (Lichtenthaler 2008).

Bresnan et al. (2003) raise the question what happens in long-term collaborations when an *"…individual leaves and takes their knowledge and contacts with them?"* (Bresnan et al. 2003, p164). We have found that where there are established, close-knit, personal networks, the departing member and their knowledge is still accessible through those networks.

In the investigated collaborative group, technology-based approaches did not play a key role in knowledge retention. This is in part due to lack of support for implementation and the members own disinterest. Instead they have developed their knowledge retention through a personalised strategy with a human-based approach through their networks (Hansen et al 1999). Our findings support the perspective that, even with the availability of many technology-based options, people are the key element of knowledge management (Swan et al 1999).

But this brings up the question, whether different technology-based approaches would further support what was happening already within EnviroAlliance and whether they would need them? Possibly they would, as with the current processes, new members especially were disadvantaged in developing background knowledge on the previous decisions and actions of EnviroAlliance. Research has demonstrated that groups with good knowledge repositories are more equipped to train new members and effectively utilise them (Kush et al. 2012). The addition of technology-based repositories accessible to all members could at least help the integration of new members in EnviroAlliance if not support their and the other group members knowledge related activities in general. Further improving knowledge sharing and retention might also extend to continuous members of the group.

# 7 Conclusion

With a focus on knowledge retention in loosely-coupled inter-organisational government-industry collaborations and utilising a combination of interviews, observations and SNA, we examined a regional government-industry collaboration focused on environmental and sustainability issues.

Our research provides several contributions. Knowledge sharing channels offer different depths of knowledge for members. This ensures that members are kept up-to-date on all activities. More importantly, group interactions also promote further development of personal networks amongst members (O'Leary et al. 2010). The end of a member's participation in a group is not the end of their knowledge sharing and retention in the group. The group can maintain contact through personal networks with these departing members. The human-based approach utilising personal networks allows a group's knowledge retention where technology-based approaches are unavailable (Hansen et al. 1999). However, technology-based knowledge retention approaches could provide additional support in particular for new members with a more effective integration into the group. As our findings are limited to analysis of one government-industry collaboration, further research is needed to explore the balance between human- and technology-based approaches. However, the reflection of our findings in the existing theory on collaboration and knowledge retention provides generalisation beyond the previously researched domains (Lee and Baskerville 2003).

Finally, we also demonstrate the potency of SNA which aids in identifying the strength and continued contact of relationships in the personal networks. SNA proved to be an invaluable research method and provided valuable insights as the findings from the SNA were not evident through observation and interviews but emerged through analysis of the personal networks.

## Acknowledgements

We would like to acknowledge the contributions of A/Prof Annemieke Craig and A/Prof Jamie Mustard in development of this work.

## Copyright